# THE RISK ASSESSMENT AND TREATMENT APPROACH IN ORDER TO PROVIDE LAN SECURITY BASED ON ISMS STANDARD

Marzieh Sameni Toosarvandani [1] and Nasser Modiri [2] and Mahdi Afzali[3]

[1]Department of Computer Engineering, Zanjan Branch, Islamic Azad University, Zanjan, Iran
m.sameni14@yahoo.com

[2]Department of Computer Engineering, Zanjan Branch, Islamic Azad University, Zanjan, Iran
nassermodiri@yahoo.com

[3]Department of Computer Engineering, Zanjan Branch, Islamic Azad University, Zanjan, Iran
afzali@hacettepe.edu.tr

## ABSTRACT

*Local Area Networks(LAN) at present become an important instrument for organizing of process and information communication in an organization. They provides important purposes such as association of large amount of data, hardware and software resources and expanding of optimum communications. Becase these network do work with valuable information, the problem of security providing is an important issue in organization. So, the stablishment of an information security management system(ISMS) in organization is significant. In this paper, we introduce ISMS and its implementation in LAN scop. The assets of LAN and threats and vulnerabilities of these assets are identified, the risks are evaluated and techniques to reduce them and at result security establishment of the network is expressed.*

## 1. INTRODUCTION

Today, by daily development of computer networks and increasment of data that exchange, security becomes an important issue for organization. Security in organization is use of strategy for getting a situation that managers could protect informations and their communications against different risks, damages and incidents that threats the organization. By increasment of using new technologies in information and communication, establishing of a ISMS in organization is essential. This system is a systematic way for designing, executing, performing, and review of information security based on standard methods[1].

The first standard of ISMS by the name of BS7999 published by business and industry govermented organization of England that consist of some parts. The first part of that is aboute information security management that published in 1995, in 1998 it revised and in 2000 it presented by international standard organization (ISO) by the name of ISO/IEC 17799. The second part of BS7999 standard prepared by coordination between this standsrd and ISO management standards in 2002. This standard in 2005 published by ISO by the name of ISO/IEC 27002. By this standards, establishment of information security in organization should be continual and in protection making cycle that it´s stages are designing , executing, assessment and correction[2]. From that date, many standards of ISO/IEC 27000 series published or thay are making ready. In this paper, we introduce steps of establishing ISMS(part 2), standards, domains and implementation model of ISMS process(part 3,4,5), The risks in LAN and Selecting of appropriate controls and allocating mitigation technique them (part 6) and at last the result of this paper will be present(part 7).

## 2. STEPS OF STABLISHING ISMS

ISMS is an approach to protecting and managing data based on a systematic method to establish, implement, operate, monitor, review, maintain, and improve of information security[3]. The steps of establishing ISMS explained as: Define the scope, planning ISMS policies, risk assessment, risk treatment, selecting the management controls and preparing statement of applicability.

### 2.1. Define the Scope

Define the scope of ISMS can be implemented for just a department, or the entire or part of an organization. Selecting of this scope involves several factor, as: the information security policy, the information security objectives and plans, the roles and responsibilities that are related to information security and are defined by the management[4]. We have considered the LAN and its devices for the ISMS scop.

### 2.2. Planning ISMS Policies

ISMS policies set out the basic concepts for  information management in a organization. The policies must state the general focus of information security and act as guide to action. The points in procedure of  planning ISMS policies are as: establishing ISMS policies, structure of organization for establishing the ISMS, endorsement from management[5]. We have considered LANs assets management and devices security according to the organization's policies, in this paper.

### 2.3. Asset, Threat, Vulnerability and Risk Assessment

Every organization is continuously exposed to an endless number of new or changing threats and vulnerabilities that may affect its operation or the fulfillment of its objectives. Identification, analysis and evaluation of these threats and vulnerabilities are the only way to understand and measure the impact of the risk involved and hence to decide on the appropriate measures and controls to manage them[6].

### 2.3.1. Identifing Assets Value

Asset value can be defined by CIA factors that are confidentiality, integrity and availability of an asset. Confidentiality is ensuring that information is accessible only to those authorized to have access. Integrity is safeguarding the accuracy and completeness of Information and processing methods. Availability is ensuring that authorized users have access to information and associated assets when required[7]. Various methods are also available for finding assets value, one of these methods is to the assets are evaluated in terms of the influence of confidentiality, availability and integrity factors and values are added together, as (1)[8]:

(1) Asset value = Confidentiality value + Availability value + Integrity value

Each of these factors can be set from low, medium and high, respectively, with the numbers 1 to 3 are based on their value. Oure assets in LAN are devices in network such as router and switches.

### 2.3.2. Identifing Threats and Threats Value

A threat is a potential cause of a security incidents that may cause an information system or organization to be lost or damaged. Threats are caused by vulnerabilities and efect the organization of the operations or the organization if they occur. After identifying threats related to network layer, according to probability of occurring threats, we allocate values of low, medium and high, respectively, with the numbers 1 to 3 for each of threat[5].

### 2.3.3. Identifing Vulnerabilities and Vulnerabilities Value

Vulnerabilities are weak points and security holes that may cause threats, and which are specific to an asset. Vulnerabilities do not cause any damage by themselves, but they may help threats to occure or cause damages[5]. After identifying vulnerabilities associated with each threat, the value of each vulnerability according to vulnerability impacts on threat, is determined. We use for it valsue of low, medium and high, respectively, with the numbers 1 to 3. thraets by using vulnerabilities, creates risks[9]. Figure 1 shows this cycle.

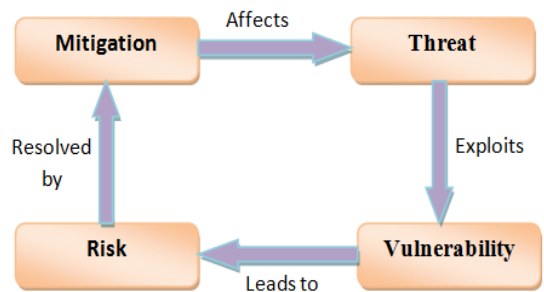

Figure 1. The ralation between threats, vulnerabilities and risks

### 2.3.4. Risk Assessment

Risk assessment can be performed by determining assessment procedures, creating an inventory of assets and clarifying the classificatins of impotance of assets and the crifteriafor evalution threats, vulnerabilities and risks. We use the formula to calculate risk values, as (2).

(2) Risk value=value of asset*value of threat* value of vulnerability

Then according to the level of risk acceptance in organization, for risks that are above this level, we will decide during the risk management process[5].

### 2.4. Risk Treatment

Risk treatment is the process of selecting and implementing of measures to modify risk. These measures can be selected out of sets of security measurements that are used within the ISMS of the organization[6]. Risk treatment have different methods of handling risks, as: risk acceptance, risk avoidance, risk avoidance and risk transfer.

### 2.4.1. Risk Acceptance

To accept the risk and continue operating or to implement controls to lower the risk to an acceptable level[8]. Selecting a set of risk acceptance criteria based on the goals and objectives of the organization is important to have as an integral part of the ISMS. This assists in the development of risk treatment plans[10].

### 2.4.2. Risk Avoidance

Risk avoidance means performing methods to consider the risk treatment and abolishing an operation or discarding information assets to avoid a risk if it is impossible to prevent it adequately for a reasonable cost or no appropriate measures for dealing with it are found[5].

### 2.4.3. Risk Limitation

In the real world, risks cannot be completely removed by taking measures. In many cases, a measure is performed by investing enough to keep any risks which may occure under the acceptable level[5]. To limit the risk by implementing controls that minimizes the adverse impact of a threat's on an asset[8].

### 2.4.4. Risk Transfer

Risk transfer means transferring risks to other partties or other organizations by concluding a contract. Risk transfer can be divided into two main categories: outsourcing measures for information assets and information security, and using an insurance system as a kind of finance for risks[5].

## 2.5. Selecting the Management Goals and Controls

The method of using an appropriate control to limiting the risk is the most common form of risk management. In this case, risk can be reduced by selecting a management goals and controls from the ISMS standard. Useful controls can be from existing controls or mechanisms that are in standards and  guide directions of information security or they are from suggested control for organizational needs or special operations characteristics.

## 2.6. Preparing Statement of Applicability

This step is ready the documentation of all works we had been done, as document the management goal and control selected in step 5, and the reason for choosing them and prepare a Statement of Applicability[6].

## 3. ISMS STANDARDS

ISO provides several documents that offer guidance in developing the ISMS. Those relevant to management of risk are:

### 3.1. ISO/IEC 27001

This standard is related to requirements of ISMS. It describes a model for establishing, implementing, operating, monitoring, reviewing, maintaining, and improving an ISMS. It used to assess conformance by interested internal and external parties and applies to all types of organizations and ensures selection of adequate and proportionate security controls that protect information assets and give confidence to interested parties. This standard specifies requirements for the implementation of security controls customized to the needs of individual organizations or departments. It provides 11 domains of information security[10].

### 3.2. ISO/IEC 27002

This standard is related to code of practice for ISMS. It defines security controls that may be selected within each domain of ISO/IEC 27001 and provides implementation guidance in each area[10].

## 4. ISMS DOMAINS

ISO/IEC 27001 standard has 11 domains[8], which these domains are listed in table 1. In this paper, according to the objectives, policies and requirements of the organization, we have considered asset management, physical and environmental security, human resource security, Information Security Incident Management domains of ISMS.

Table 1.  ISMS domains

| No. | Domain name | Description |
| --- | --- | --- |
| 1 | Security Policy | Management should approve the information security policy, assign security roles and co-ordinate and review the implementation of security across the organization |
| 2 | Organizing Information Security | Infrastructure, third party access and controlling security of outsourced information processing |
| 3 | Asset Management | Identifying, classifying and protecting assets and information |
| 4 | Human Resource Security | Addressing roles and responsibilities, screening, training, disciplinary process, termination |
| 5 | Physical and Environmental Security | Managing physical access to prevent loss, damage, theft, compromise |
| 6 | Communications and Operations Management | Ensuring correct and secure operations in computer and network systems, third party services, media, electronic messaging, monitoring |
| 7 | Access Control | Controlling access to information, enforced by controlling and monitoring access rights to networked devices, operating systems, applications, both directly on the organization's network and via remote access |
| 8 | Information Systems Acquisition, Development and Maintenance | Building security into the information systems |
| 9 | Information Security Incident Management | Damage control, reporting, collecting evidence and root cause for the incident. Preventive action |
| 10 | Business Continuity Management | Counteracting interruptions and minimizing their impact |
| 11 | Compliance | Avoiding breaches of law, regulatory or contractual requirements |

## 5. IMPLEMENTATION OF ISMS PROCESS IN PDCA MODEL

PDCA cycle presented by Edward Deming, was a model for making better the procedure or system that is in figure 2. This cycle is a dynamic model, in a way that the last step of evolution from previous stage, is the first step in next stage and by repetition of procedure improvement can be analyzed always[11]. This cycle has 4 steps, In Plan step, we define needs, evaluate risks and assigns executable controls. In Do step, development, use and execution of controls has been done. In Check step, inspecting the system operation done by regarding the needs, and in Act step, we pay attention to maintenance, correction and making better of system[12].Figure 2 shows PDCA model.

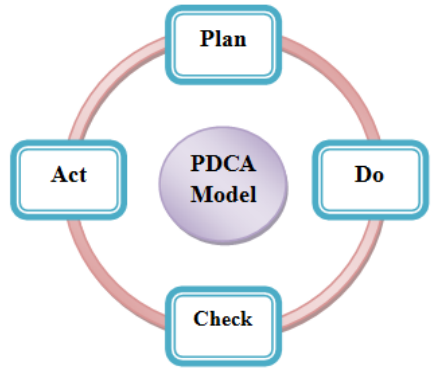

Figure 2. PDCA model

## 6. SELECTING APPROPRIATE CONTROLS FOR THREATS AND ALLOCATING TECHNIQUE THEM

In this part, we explain the approach of mapping threats to ISO/IEC 27001 standard controls, then mitigation approaches for threats of LAN based on Cisco configuration is explained. We use the eliminating risks method and reducing risks method to treatment the identified risks, by selecting control or controls from a list of controls available in ISO/IEC 27001 and their practical code from ISO/IEC 27002 for each threat and risk[13,14,15,16]. By allocaiting the appropriate control, techniques to reduce the risks and threats are expressed. Table 2 shows threats and vulnerabilities and risks of Cisco IOS software in LAN, appropriate contlols for threats and mitigation techniques of These threats. Table 3 shows above subjects for Layer 3, Table 4 shows them for Layer 2 and Table 4 shows these subjects for router and switch device.

Table 2. Threats and vulnerabilities and risks of Cisco IOS software in LAN, appropriate contlols for threats and mitigation techniques of These threats

| Asset | Threat | Vulnerability | Selected Contlol | Implementation Technique |
|-------|--------|---------------|------------------|--------------------------|
| Cisco IOS software | Memory shortage | Memory flood | 7.84.Have you implemented rules that define the acceptable use of assets associated with information | Configure the router with the maximum amount of memory possible |

| | | | processing facilities?[13] | |
|---|---|---|---|---|
| | Work restrictios for future | Using of Inefficient operating system versions | 7.84.Have you implemented rules that define the acceptable use of assets associated with information processing facilities? | Use the latest stable version of the operating system that meets the feature requirements of the network |
| | Loss of operating systems | Operating system failures | 7.84.Have you implemented rules that define the acceptable use of assets associated with information processing facilities? | Keep a secure copy of the router operating system image and router configuration file as a backup |
| | Unauthorized access | Potential abuse of unused ports and services | 7.2.Do you use controls to protect your assets? 7.22. Have you identified levels of protection for your assets? | Secure administrative control, ensure that only authorized personnel have access and that their level of access is controlled |
| | | | | Disable unused ports and interfaces, reduce the number of ways a device can be accessed |
| | | | | Disable unnecessary services, similar to many computers, a router has services that are enabled by default[17] |
| | | Lake of securing administrative access | 7.71.Are your asset owners responsible for defining access classifications and ensuring that these classifications are consistent with access control policies? 7.72.Are your asset owners responsible for reviewing access classifications and ensuring that these classifications are consistent with access control policies? 7.102.Have you specified how much protection is expected at each level? 7.117.Are your | Restrict device accessibility: Limit the accessible ports, restrict the permitted communicators, and restrict the permitted methods of access |
| | | | | Creating log and account for all accesses: For auditing purposes, record anyone who accesses a device, including what occurs and when[17]. |
| | | | | Authenticate access: Ensure that access is granted only to authenticated users, groups, and services. Limit the number of failed login attempts and the time between |

| | | | information classification guidelines consistent with your access control policy? 9.49.Do your physical entry controls allow only authorized personnel to gain access to secure areas?[15] | logins[17]. |
|---|---|---|---|---|
| | | | | Authorize actions: Restrict the actions and views permitted by any particular user, group, orservice |
| | | | | Present Legal Notification: Display a legal notice, developed in conjunction with company legal counsel, for interactive sessions |
| | | | | Ensure the confidentiality of data: Protect locally stored sensitive data from viewing and copying[17] |
| | | Lake of security Remote access such as Telnet, SSH, HTTP, HTTPS, or SNMP connections to the router from a computer | 7.23.Have you assigned a level of protection to each asset? 7.24.Do you provide a higher level of protection for your organization's most valuable and important assets? | Encrypt all traffic between the administrator computer and the router, For example, instead of using Telnet, use SSH[17]. |
| | | | | Establish a dedicated management network. The management network should include only identified administration hosts and connections to a dedicated interface on the router[17]. |
| | | | | Configure a packet filtering to allow only the identified administration hosts and preferred protocols to access the router[17]. For example, permit only SSH requests from the IP address of the administration host to initiate a connection to the routers in the network[17] |

| | | Lake of Security for Virtual Logins | 7.79.Have you identified rules that define the acceptable use of information? 7.81.Have you implemented rules that define the acceptable use of information? | Delays between successive login attempts |
|---|---|---|---|---|
| | | | | Login shutdown if DoS attacks are suspected |
| | | | | Generation of system logging messages for login detection[18] |
| | | Do not use passwords for all lines and terminals | 7.82.Have you identified rules that define the acceptable use of assets associated with your information processing facilities? 7.99.Do you provide an appropriate level of protection for your organization's information? | Enable Secret Password: It restricts access to privileged EXEC mode. The enable secret password is always hashed inside the router configuration using a Message Digest 5 (MD5) hashing algorithm[17]. If the enable secret password is lost or forgotten, it must be replaced using the Cisco router password recovery procedure |
| | | | | Creating Password for Console Line: By default, the console port does not require a password for console administrative access, it should always be configured as a console port line-level password |
| | | | | Creating Password for Virtual Terminal Lines: By default, Cisco routers support up to 5 simultaneous virtual terminal vty (Telnet or SSH) sessions. On the router, the vty ports are numbered from 0 through 4[17] |
| | | | | Creating Password for Auxiliary Line: By default, Cisco router auxiliary ports do not require a password for remote administrative access. Administrators sometimes use this port to |
| | | | | remotely configure and monitor the router using a dialup modem connection[17] |
| | | Selecting of weak passwords to protect assets | 7.24.Do you provide a higher level of protection for your organization's most valuable and important assets? | Use a password length of 10 or more characters[17]. |
| | | | | Make passwords complex. Include a mix of uppercase and lowercase letters, numbers, symbols, and spaces |
| | | | | Avoid passwords based on repetition, dictionary words, letter or number sequences, usernames, relative or pet names, biographical information or other easily identifiable pieces of information |
| | | | | Deliberately misspell a password. For example, Security = 5ecur1ty |
| | | | | Change passwords often. If a password is unknowingly compromised, the window of opportunity for the attacker to use the password is limited |
| | | | | Do not write passwords down and leave them in obvious places such as on the desk or monitor |
| | | | | Minimum Character Length: It is strongly recommended that the minimum password length be set to at least 10 characters to eliminate common passwords that are short and prevalent[17] |
| | | | | Disable Unattended Connections: By default, an administrative interface stays active and logged in for 10 minutes after the last session activity. If an administrator is away |

| | | | | from the terminal while the console connection is active, an attacker has up to 10 minutes to gain privilege level access. It is recommended that these timers be finetuned to limit the amount of time to within a two or three minute maximum[17] |
| | | | | Encrypt All Passwords: By default, some passwords are shown in plain text, meaning not encrypted, in the Cisco IOS software configuration. With the exception of the enable secret password, all other plain text passwords in the configuration file can be encrypted in the configuration file[17] |
| | Don't Assigning Administrative Roles | Don't Assigning Privilege Levels Assigning Privilege Levels: Privilege levels determine who should be allowed to connect to the device and what that person should be able to do with it | 9.130.Do you isolate your equipment when it requires an extra level of protection? | User EXEC mode (privilege level 1): Provides the lowest EXEC mode user privileges and allows only user-level commands |
| | | | | Privileged EXEC mode (privilege level 15): Includes all enable-level commands[17] |
| | Lake of monitoring and management | Lake of logging for all accesses | 7.70.Do you make your asset owners responsible for protecting your organization's assets even though owners may have delegated the responsibility for implementing controls? 7.130.Have you | The Cisco IOS Resilient Configuration feature: This feature allows for faster recovery if someone reformats flash memory or erases the startup configuration file in NVRAM. This feature allows a router to withstand malicious attempts at erasing the |

| | | | developed information handling procedures for each of your information security classifications? 8.31.Do your security roles and responsibilities make it clear that security risks must be reported to your organization?[14] | files by securing the router image and maintaining a secure working copy of the running configuration[17] |
| --- | --- | --- | --- | --- |
| | | | | Implementing a router logging to Console: Implementing a router logging is an important part of any network security policy. Cisco routers can send log messages to several different facilities. Console logging is on by default Messages log to the console and can be viewed when modifying or testing the router using terminal emulation software while connected to the console port of the router[17]. |
| | | | | Implementing a router logging to Terminal lines: Enabled EXEC sessions can be configured to receive log messages on any terminal lines. Similar to console logging, this type of logging is not stored by the router and it is only valuable to the user on that line |
| | | | | Buffered logging: Buffered logging is a little more useful as a security tool because log messages are stored in router memory for a time[17]. The events are cleared whenever the router is rebooted |
| | | | | SNMP traps: Certain thresholds can be preconfigured on routers and other devices. Router events, such as exceeding a threshold, can be processed by the router and forwarded as SNMP |

| | | | | traps to an external SNMP server. SNMP traps are a viable security logging facility but require the configuration and maintenance of an SNMP system[17] |
|---|---|---|---|---|
| | | | | Syslog service: Cisco routers can be configured to forward log messages to an external syslog service. Syslog is the most popular message logging facility, because it provides long term log storage capabilities and a central location for all router messages[17] |
| | Lake of Security Audit | Exploiting from vulnerable points that will be create on future | 7.3. Do you account for your organization's assets? 7.132.Have you implemented your information handling procedures? 13.14.Establish a formal information security event reporting procedure.[16] | Disable unnecessary services and interfaces |
| | | | | Disable and restrict commonly configured management services, such as SNMP |
| | | | | Disable probes and scans, such as ICMP |
| | | | | Ensure terminal access security |
| | | | | Disable gratuitous and proxy Address Resolution Protocol (ARP)[19] |
| | | | | Disable IP directed broadcasts |
| | | | | Security Audit Wizard: It provides a list of vulnerabilities and allows the administrator to choose which potential security related configuration changes to implement on a router[17]. |
| | | | | Cisco AutoSecure: This feature initiates a security audit and allows for configuration changes. Based on the mode selected, configuration changes can be automatic or require network administrator input[17] |

27

| | | | | One Step Lockdown: This feature provides a list of vulnerabilities and then automatically makes all recommended security-related configuration changes[17] |
|---|---|---|---|---|

Table 3. Threats and vulnerabilities and risks of Layer 3 in LAN, appropriate contlols for threats and mitigation techniques of These threats

| Asset | Threat | Vulnerability | Selected Contlol | Implementation Technique |
|---|---|---|---|---|
| Layer 3 | Denial of Service (DoS) Attack: It implies that an attacker disables or corrupts networks, systems, or services with the intent to deny services to intended users[20] | SYN Flood Attack: It sends TCP connections requests faster than a machine can process them. In SYN flood attack, an attacker creates a random source address for each packet[21] | 7.100.Have you established an information classification system for your organization? 7.101.Do you use your classification system to define security levels? 7.107.Do you classify information according to how sensitive it is? 7.108.Do you classify information according to how critical it is? 7.143.Do your information handling procedures define how information should be processed at each Security classification level? | Using an ACL to Characterize SYN: Use the ACL to disply the access list packet match statistics to identify the SYN attack |
| | | | | TCP Intercept: TCP Intercept feauture works by intercepting and validating all incoming TCP connection requests flowing between a TCP client and TCP server |
| | | | | CBAC: Context Based Access Control (CBAC) intelligently filters TCP and UDP packets based on application layer protocol session information[20] |
| | | | | IP Source Tracker: IP Source tracking is the process of tracing packet streams from the victim back to the point of origin to find the source of the attack through the network path |
| | | | | Packet Classification and Marking with CAR: The policing feature of Committed Access Rate (CAR) controls the maximum rate of traffic sent or received on an interface for a network specifying traffic handling policies, when the traffic either conforms to or exceeds the specified rate limits |
| | | | | Packet Classification and Marking with MQC: It provides a modular and extensible framework that allows users to create hierarchical traffic policies to deliver extremely powerful and scalable |
| | | | | solutions[23] |
| | | | | Packet Classification and Marking with NBAR: NBAR is a classification engine that can recognise a variety of applications and protocols from Layer 4 through Layers 7, including web based and other protocols that static and dynamically assigned TCP and UDP port numbers[22] |
| | | | | Traffic Policing: The Cisco IOS Traffic Policing features allow the control and filtering of the incoming and outgoing traffic rate on an interface,as well network bandwidth management through the token bucket algorithm |
| | | | | NetFlow: Net Flow captures statistics on IP packets flowing through the router and is emerging as a primary security technology. NetFlow classifies packets by the direction of their flow and identifies packet flows for both ingress and egress IP packets[23] |
| | Distribut ed Denial of Service Attack (DDoS): They are designed to saturate network links with spurious data[20] | ICMP Flood or Smurf Attack: It starts with a perpetrator sending a large number of spoofed ICMP echo, requests to broadcast addresses, hoping that these packets will be magnified and sent to the spoofed addresses[20] | 7.22.Have you identified levels of protection for your assets? | Using an ACL to Traffic Characterize Smurf: The Cisco IOS Access Control List (ACL) is the technique to classify the packets into various attack streams, and it is valuable for characterizing attacks and tracing packet streams back to their point of origin |
| | IP spoofing Attack: With a masquera | Falsifying the source IP address that is within the range of IP | 7.112.Do your information classifications allow you to meet your | Antispoofing with Access Lists: It is for droping packets that arrive on interfaces that are not viable paths from the supposed source addresses of those |

| | | | |
|---|---|---|---|
| de attack, the network intruder can manipula te TCP/IP packets by IP spoofing, falsifying the source IP address, thereby appearin g to be another user | addresses for the network or authorized external IP address that is trusted | business need to restrict access to information? 7.114.Do your protective control methods allow you to meet your business need to restrict access to information? | packets[23] |
| | | | Antispoofing with uRPF: When Unicast Reverse Path Forwarding (uRPF) is used, the source address of IP packets is checked to ensure that route back to the source uses the same interface that the packet arrived on[24] |
| | | | Antispoofing with IP Source Guard: IP Source Guard prevents IPspoofing attack by restricting IP traffic on untrusted Layer 2 ports to clients with an assigned IP address |

Table 4.  Threats and vulnerabilities and risks of Layer 2 in LAN, appropriate contlols for threats and mitigation techniques of These threats

| Asset | Threat | Vulnerability | Selected Contlol | Implementation Technique |
|---|---|---|---|---|
| Layer 2 | MAC Address Table Overflow Attack: MAC spoofing attacks occur when an attacker alters the MAC address of their host to match another known MAC address of a target host[23] | Flooding the switch by a large number of frams with invalid source and destination MAC address | 7.2.Do you use controls to protect your assets? | Port Security feature by By limiting the number of permitted MAC addresses on a port to one, port security can be used to control unauthorized expansion of the network |
| | MAC Address Spoofing Attack: bombarding the switch with fake source MAC addresses until the switch MAC address table is full[17]. If enough entries are entered into the MAC address table before older entries expire, the table fills up to the point that no new entries can be accepted[25] | Switch Spoofing by forging the MAC address of target host | 7.2.Do you use controls to protect your assets? | Port Security feature[23] |
| | ARP Spoofing Attack: The intruder can inject an | Forwarding of requested ARP | 9.6.Do you use | Configuration the hold-down timers on |

| | | | |
|---|---|---|---|
| | unsolicited fake ARP reply message with its own MAC address, sending this message to the requester first host masquerading as the victim, second host[26] | reply message with Spoofing MAC address and poisoning ARP cache | appropriate security barriers to protect your critical or sensitive information processing facilities? 9.125.Do you protect your equipment from environmental risks and hazards through the use of secure siting strategies? | the interface by specifying the duration of time for an ARP entry to remain in the ARP cache |
| | | | | To enable of Dynamic ARP Inspection(DAI) feature on switch (best technique)[23] |
| | VTP Attack: Intruder can send falsified VTP massages on a trunk port, posing as a VTP server, and gaining privilege tp add or remove VLANs from the VTP domain as well as to create Spanning Tree protocol loops[27] | Sending falsified messages on a trunk port of switch and adding or removing VLANs from VTP domain and creating STP loops | 9.7.Do you use entry controls to protect your critical or sensitive information processing facilities? | Configure a unique VTP domain name with a strong VTP password[23] |
| | VLAN Hopping Attack: A VLAN Hopping attack is the technique of jumping VLANs without traversing a Layer 3 device[20] | Spoofing of switch with emulating either ISL or 802.1Q signaling along with DTP signaling | 7.99.Do you provide an appropriate level of protection for your organization's information? 9.130.Do you isolate your equipment when it requires an extra level of protection? | Turning off DTP on all user ports except the ports that specifically require DTP,such as the trunk port. Dynamic trunking protocpl(DTP) is a Layer 2 protocpl used to automate ISL and 805.1Q trunk configurations between switches and supports autonegotiation of both ISL and 805.1Q trunks. |
| | | | | Native VLAN-ID on all the trunk ports must be different from the Native |

| | | | VLAN-ID of the user ports[23] |
|---|---|---|---|
| | PVLAN Attack: In PVLAN attack, frames are forwarded to a host that is connected to a promiscuous port, That is primarily used to defeat PVLAN Configuration by avoiding the promiscuous port[20] | Forwarding of frams to a host on the network that is connected to promiscuous port | To configure ACLs or virtual ACLs on the default gateway router to block any arriving packets that have the same source and destination IP address[23] |
| | | 7.25.Do you provide a higher level of protection for assets that have a higher security classification? | |
| | Spanning Tree Attack: Spanning Tree protocol(STP) is a link management protocol that provides path redundancy by preventing loops in a network of switches, The intruder can possibly force all the switches to forward packets to the intruder switch by injecting falsified Bridge Protocol Data Unit(BPDU) with a priority zero and forcing spanning Tree recalculations so that the intruder switch can become the new root bridge and traffic is transmitted across the attacker switch[20] | Injecting falsified BPDU with a priority zero and forcing Spanning Tree recalculations so that the intruder switch can become the new root bridge | BPDU Guard feature: If a port that is configured with PortFast(PortFast can be used on Layer 2 access ports that connect to a single workstation or server to allow those devices to connect to the network immediately, instead of waiting for STP to converge) receives a BPDU, STP can put the port into the disabled state by this feature |
| | | 9.126.Do you prevent opportunities for unauthorized access to equipment through the use of secure siting strategies? 9.128.Do you position information processing facilities so that sensitive information cannot be viewed by unauthorized persons? | ROOT Guard feature that it limits the switch ports out of which the root bridge can be negotiated, If a root guard enabled port receives BPDUs that are superior to those that the current root bridge is sending, that port is moved to a root-inconsistent state, which is effectively equal to an STP listening state, and no data traffic is forwarded across that port[17] |
| | DHCP Spoofing and | Flooding a | 7.84.Have | Port Security feature |

| | | | |
|---|---|---|---|
| | Starvation Attack: works on MAC address spoofing by flooding a large number of DHCP requests with randomly generated spoofed MAC address to the target DHCP server [28] | large number of DHCP requests with randomly generated spoofed MAC address to the target DHCP server | you implemented rules that define the acceptable use of assets associated with information processing facilities? 7.110.Have you implemented rules that define the acceptable use of assets associated with information processing facilities? | VLAN ACLs to mitigate rogue DHCP server by preventing the rogue server from responding to DHCP requests |
| | | | | To enable the DHCP snooping feacure available on switch (best technique)[23] |

Table 5.  Threats and vulnerabilities and risks of Router and switche device in LAN, appropriate contlols for threats and mitigation techniques of These threats

| Asset | Threat | Vulnerability | Selected Contlol | Implementation Technique |
|---|---|---|---|---|
| Router and switche device | Power outage | Don't use of UPS | 9.167.Do you use uninterruptible power supplies (UPSs) to protect equipment that is used to support critical business operations? 9.174.Do you test your backup generators regularly in accordance with manufacturers' recommendations? | Install an uninterruptible power supply (UPS) and keep spare components available and check out it to ensure the performance |
| | Theft | Putting tools and equipment in unsafe places | 9.117.Do you prevent the theft of your organization's equipment? 9.132.Do you use security controls to minimize the risk that equipment will be stolen? | Place the router and physical devices that connect to it in a secure locked room that is accessible only to authorized personnel |
| | Fire | Placing combustible materials near | 9.90.Is appropriate fire fighting equipment suitably situated and | Putting in place the necessary tools to extinguish the fire, avoid placing combustible |
| | | critical assets, lack of access to tools to extinguish the fire | available when needed? 9.134.Do you use security controls to minimize the risk that equipment will be damaged by fire? | tools in important places, Placing equipment according to the fire exit doors |
| | natural disasters | Improper establishment of facilities against natural disasters | 9.76.Do you use physical methods to protect your facilities from the damage that natural disasters can cause? 9.79.Do you use physical methods to protect your facilities from the damage that floods can cause? 9.80.Do you use physical methods to protect your facilities from the damage that earthquakes can cause? 9.149.Do you protect your buildings from lightning strikes? | Appropriate establishment of facilities against earthquake and flood, using of instrument against lightning strikes |
| | Unauthorized access to edge router: The edge router is the last router between the internal network and an untrusted network such as the Internet | Disable to Edge Router for Attack to network | 9.15.Do you assess your security risks and make sure that your security perimeters actually reduce your security risk? | Single Router Approach: A single router connects the protected internal LAN to the Internet. All security policies are configured on this device[17] |
| | | | | Defense in Depth Approach: The edge router acts as the first line of defense. It has a set of rules specifying which allows or denies the traffic. The second line of defense is the firewall. The firewall typically picks up where the edge router leaves off and performs additional filtering. It provides additional access control by tracking the state of the connections and acts as a checkpoint device[17] |
| | | | | DMZ Approach: The DMZ can be set up between two routers, with an internal router connecting to the protected network and an external router connecting to the unprotected network. The firewall, located |

| | | | | between the protected and unprotected networks, is set up to permit the required connections from the outside networks to the public servers in the DMZ[17] |
|---|---|---|---|---|

# 7. CONCLUSIONS

By expanding the use of LAN and facilities and equipments of these networks in organizations, creation of security is very significant. ISMS has an important role in assigning organization´s security controls that is based on policies, aims and organizational needs. In this paper, for implementing the LAN based on the ISMS standard, assets and threats were identified in LAN and vulnerabilities and risks that threat uses them were explained. In the end by explaining the method of mapping threats to the ISMS controls, reducing techniques for these threats were expressed. By knowing these techniques and creating the statement of applicability, Plan step is finished. In Do step, these controls is implemented, this work is planned and security incidents identified. In Check step, the requirements documented, the procedures monitored and evaluated and in case any weakness .In Act step ,these weakness is identificated and effective action to prevent or correct it done. By starting the PDCA cycle again, these steps are repeated until to achieve an acceptable level of security.